\documentstyle[aps,twocolumn,eqsecnum]{revtex}

\begin{document}
\author{Itzhak Bars\thanks{%
Research supported in part by the DOE Grant No. DE-FG03-84ER-40168. }\medskip}
\title{{\small \noindent hep-th/9503205 \hfill USC-95/HEP-B1 } \bigskip\\
Ghost-free spectrum of a quantum string \\
in SL(2,R) curved spacetime}
\address{Department of Physics and Astronomy\\
University of Southern California\\
Los Angeles, CA 90089-0484}
\date{March 10, 1995\ }
\maketitle

\begin{abstract}
\centerline{\bf ABSTRACT}

\smallskip\

The unitarity problem in curved spacetime is solved for the string described
by the SL(2,R) WZW model. The spectrum is computed exactly and demonstrated
to be ghost-free. The new features include (i) SL(2,R) left/right symmetry
currents that have logarithmic cuts on the world sheet but that satisfy the
usual local operator products or commutation rules, (ii) physical states
consistent with the monodromy condition of closed strings despite the
logarithmic singularity in the currents, and (iii) a new free boson
realization for these currents which render the SL(2,R) WZW model completely
solvable.
\end{abstract}


\section{Introduction}

A string propagating in curved spacetime with one time and $(d-1)$ space
coordinates is described by an action in the conformal gauge that has the
form
\begin{equation}
S=\int d\tau d\sigma \,[\partial _{+}X^\mu \partial _{-}X^\nu \,G_{\mu \nu
}(X)+\cdots ]  \label{action}
\end{equation}
where $G_{\mu \nu }(X)$ is a background metric in $d$-dimensions with
signature $(-1,1,1,\cdots ).$ The terms in the action denoted by $\cdots $
may contain additional background fields such as an antisymmetric tensor $%
B_{\mu \nu }(X)$ a dilaton $\Phi (X)$ etc.. The overall theory must be
conformally invariant at the quantum mechanical level. In general it is
difficult to impose the conformal invariance condition non-perturbatively,
but some solutions do exist. However it is even more difficult to solve the
system. Thus, up to this point there has not been any solution presented to
the quantum mechanical spectrum or other quantum mechanical features of a
string propagating in curved space and time. There are some solutions of
models for 2D gravity. But the spacetime interpretation remains somewhat
obscure in the matrix formulation of such models and furthermore there seems
to be no prospect for extending the approach to higher dimensions.

One knows the spectrum and correlation functions for many string models in
curved space (without time) in several dimensions; these are the string
models for compactified space that represent possible string vacuua. However
when the time coordinate is included as part of the curved space, as it
would be during the early universe, little is known about the quantum string
theory. The lack of such solutions has prevented the understanding of the
role and true impact of string theory on the structure of symmetries and
matter as observed in todays universe. If string theory is truly relevant,
it must play its major role in the presence of quantum gravity during the
early part of the universe when spacetime is curved. Thus, the structure of
gauge symmetries, matter content in the form of families of quarks and
leptons, and all that, would be determined by the structure of string theory
while the universe is at an early stage with its time coordinate being part
of the curved spacetime. Given this intuitive fact, it would be premature to
try to predict from string theory the structure of low energy physics (at
accelerator energies) by considering only flat four dimensional spacetime
plus additional curved compactified spaces in extra dimensions, as is
usually done by string phenomenologists. The role of quantum gravity and its
impact on low energy physics could not be assessed without a better
understanding of strings during the early universe while time and space are
curved.

Many models that are exactly conformally invariant in one time plus $(d-1)$
space dimensions have been constructed by now \cite{BN}\cite{ibcurved}\cite
{tseytlinrev}. But no further undersdanding of the quantum properties of
these models has been obtained. One of the stumbling blocks has been the
issue of the unitarity of such theories (negative norm states) \cite{BN}\cite
{O'R}. Although there has been many attempts to crack the problem \cite
{ibstonybrook}\cite{attempts}, no real dent has been made until now.

In this paper we solve the unitarity problem that has plagued a certain
class of models, and provide new free field methods for the computation of
many quantum properties of the models. In particular we give the explicit
solution of the exact spectrum of a string propagating in the SL(2,R) curved
spacetime of a WZW model. This is the simplest of all curved spacetime
models. This resolves also the unitarity and spectrum issues for the
SL(2,R)/R black hole model \cite{BN}\cite{WIT} and provide the methods of
spectrum computation for all higher dimensional models that are in a similar
class \cite{BN}\cite{ibcurved}, including supersymmetric and heterotic
versions. The free field methods that are introduced in this paper open the
way to the computation of correlation functions as well.

\section{The unitarity problem}

One of the first attempts to solve string propagation in curved spacetime
was to consider some current algebra models that potentially can be solved
through algebraic methods \cite{ibcurved}. However, one immediately comes
accross an unexpected problem with the unitarity of the theory. Due to the
presence of a time coordinate there are negative norm oscillators. These
create negative norm states. However, since one must also impose the
Virasoro constraints, on the basis of na\"\i ve counting one may hope to
prove that all negative norm states are removed from the theory. A similar
situation occurs also in the flat theory. As is well known, in the flat case
one can indeed prove the no ghost theorem \cite{noghost} which implies that
the theory is unitary. However, in the case of curved spacetime current
algebra models, one finds that even after imposing the Virasoro constaints
there remains negative norm states that render the theory non-unitary. This
has been the main stumbling block that discouraged the application of these
ideas to model building for the past five years.

To illustrate the problem consider SL(2,R) currents, and a stress tensor
defined in the form of Laurent series
\begin{eqnarray}
\tilde J^i(z) &\equiv &\sum_{n=-\infty }^\infty J_n^i\,z^{-n-1},\quad
i=0,1,2,  \nonumber \\
\tilde T(z) &\equiv &\sum_{n=-\infty }^\infty \tilde L_n\,z^{-n-2}
\label{laurent} \\
\tilde L_n &\equiv &\frac 1{k-2}\sum_{m=-\infty }^\infty
:J_{-m}^iJ_{n+m}^j:\,\eta _{ij}  \nonumber
\end{eqnarray}
where $\eta _{ij}=diag(-1,1,1)$ is the Minkowski metric and is proportional
to the Killing metric for SL(2,R). The commutation rules are
\begin{eqnarray}
\left[ J_n^i,J_m^j\right] &=&i\epsilon ^{ijk}\eta _{kl}J_{n+m}^l+\frac k2%
n\delta _{n+m}\eta ^{ij}  \nonumber \\
\left[ \tilde L_n,J_m^i\right] &=&-m\,J_{n+m}^i  \label{commutators} \\
\left[ \tilde L_n,\tilde L_m\right] &=&\left( n-m\right) \,\tilde L_{n+m}+%
\frac c{12}n\left( n^2-1\right) \,\delta _{n+m}  \nonumber \\
c &=&3k/(k-2).  \nonumber
\end{eqnarray}
$J^0$ corresponds to the compact generator, and the other two correspond to
the non-compact generators. For positive $k$ these commutation rules imply
that there is one time and 2 space dimensions. One indication of this is the
large $k$ limit (semi-classical limit) in which the commutation rules of the
currents reduce to those of flat spacetime oscillators, with one time-like
and two space-like oscillators. Therefore, it is useful to think of $J^0$ as
a current in the time-like direction and of $J^{1,2}$ as two currents in two
space-like directions. Since we are interested in the covering group of
SL(2,R), the time coordinate can take all values, and it is not restricted
to a compact range. For example, one can see this by analyzing the classical
solutions of either the particle or the string in the SL(2,R) curved
spacetime in the form of a WZW model. In the large $k\rightarrow \infty $
limit the model reduces to flat spacetime in 2+1 dimensions with $c=3$. But
for finite $k$ it describes curved spacetime with a metric induced by the
group, as seen in the WZW model formulation (see section VII), with $%
c=3k/(k-2),$ which can be made critical $c=26$ if desired. Otherwise, the $%
c<26$ theory may be considered a piece of a critical theory.

The states of the theory are constructed as usual from the current algebra
Verma module (analog of Fock space of flat spacetime)
\begin{equation}
|\psi >=\prod \left( J_{-n}^i\right) ^{p_{in}}\,|jm>
\end{equation}
where $|jm>$ is a representation of the zero mode currents. The physical
states $|\phi >$ are those linear combinations of Verma module states that
satisfy the Virasoro constraints
\begin{eqnarray}
\tilde L_n|\phi &>&=0,\rm{ }\quad n\geq 1  \nonumber \\
\tilde L_0|\phi &>&=\tilde a\,|\phi >.
\end{eqnarray}
In a critical theory one may take $\tilde a=1$ and $c=26$ (or $k-2=6/23).$
It is also possible to take the SL(2,R) model as a piece of a critical
conformal theory. Then $\tilde a\leq 1$ and $c\leq 26$ (or $k-2\geq 6/23).$

The eigenvalues of $\tilde L_0$ are determined by the Casimir and the level
of excitation of the string $l\equiv \sum_{i,n}np_{in}$%
\begin{equation}
\tilde L_0=\frac{-j(j+1)}{k-2}+l=\tilde a  \label{balanc}
\end{equation}
Thus, an excited string at level $l\geq 1$ must have positive values of $%
j(j+1)$ and $j+1$ that depend on the level of excitation\footnote{%
The analog of this equation in flat spacetime is the mass shell condition $-
$($p_\mu )^2=M^2=l-\tilde a\,$ where $|p_\mu >$ denotes the labelling of the
base.}
\begin{eqnarray}
j(j+1) &=&(k-2)(l-\tilde a)  \nonumber \\
j+1 &=&\frac 12+\sqrt{(k-2)(l-\tilde a)+1/4}.  \label{jl}
\end{eqnarray}
Such values of the Casimir can occur in unitary representations of SL(2,R)
only for the discrete series $|jm>$, for which the lowest (or highest) state
has
\mbox{$\vert$}
$m|=j+1>0,$ and then the values of $m$ are given by
\mbox{$\vert$}
$m|=(j+1+n),$ where $n$ is a positive integer.

It is easy to see that there are plenty of excited states that satisfy the
Virasoro constraint, but whose norm is negative. An explicit example is \cite
{BN}
\begin{eqnarray}
\begin{array}{l}
|\phi ,l>=\left( J_{-1}^1-iJ_{-1}^2\right) ^l\,|j,m=j+1>, \\
\tilde L_n|\phi ,l>=0,\quad \quad n\geq 1, \\
<\phi ,l|\phi ,l>=N_{j(l)}\,\left( l!\right) \prod_{r=0}^{l-1}\left(
k-2j(l)-2+r\right) .
\end{array}
\label{negnorm}
\end{eqnarray}
where $N_{j(l)}=<j,m=j+1\,|j,m=j+1>$ is the norm of the state at the base.
Evidently, for sufficiently large values of the excitation number $l$ the
norm switches between positive and negative values. Hence, despite the
Virasoro constraints this model is not unitary and cannot describe a
physical string.

Until now a solution to this problem, and the related SL(2,R)/R black hole
problem, has not been found despite many attempts\footnote{%
There has been a claim of a proof of the no-ghost theorem for this problem
(second reference in \cite{attempts}). Evidently the ``proof'' is wrong
since we are able to display explicitly negative norm states that satisfy
the Virasoro conditions.} \cite{ibstonybrook}\cite{attempts}. Suggestions
included: (1) Restrict (artificially) $j(l)+1<k/2$ so that the norm never
becomes negative; (2) Allow large values of $j(l)$ as needed by the excited
level $l,\,$ but also permit the base state to have negative norm $N_{j(l)}$
in such a way as to make the norm of the excited state $<\phi ,l|\phi ,l>$
positive; (3) Hope that modular invariants will fix the problem.

All of these suggestions are rejected as follows. (1) If $j+1$ cannot exceed
$k/2$ then the string cannot be excited to arbitrary levels, as seen from (%
\ref{jl}). In addition to being an artificial condition not justified by the
formalism, this also leads to inconsistent physical results: for example,
classical string solutions in which the string is arbitrarily excited exist
in curved spacetime, there is no intuitive or physical reason not to expect
them in the quantum theory as well. (2) Even if the explicit state in (\ref
{negnorm}) is forced to have positive norm by changing the norm of the base,
there are other states at the same level that would have opposite norm to $%
|\phi ,l>$. For example, the state $|\phi ,l,+>\equiv (J_0^1+iJ_0^2)|\phi
,l>,$ which also satisfies the Virasoro constraint, has norm
\begin{equation}
<\phi ,l,+|\phi ,l,+>=2(j+1-l)<\phi ,l|\phi ,l>.
\end{equation}
The two norms have opposite signs when $j(l)+1-l$ is negative at
sufficiently high level $l$. Thus, the second suggestion does not work
either. Independent of this argument, it seems unreasonable to have a
negative norm base in a unitary theory.

The third possibility is more elusive since modular invariants are not well
understood for SL(2,R) or other non-compact groups. For the past five years
this possibility remained open. One could have hoped that the problem could
be resolved through modular invariants that are needed in order to complete
the construction of the physical theory. A modular invariant provides
instructions for putting together the left moving and the right moving
states
\begin{equation}
|\psi >=\sum \gamma _{ab}{\,}|\phi _L^a>|\phi _R^b>.  \label{modular}
\end{equation}
It could happen that the physical modular invariant would choose only those
combinations of states that have overall positive norm, even though the left
or right moving states $|\phi _L^a>,|\phi _R^b>$ contained in it may have
negative norm. However, as described below, we recently found an argument
that destroys this possibility too.

Thus, consider an open string rather than a closed one. The boundary
conditions turn out to relate the left and right moving currents, so that
only one set of currents and states are sufficient to describe the full open
string (see section IIIVc). The convenient current is neither the left nor
the right mover, but it can be related to either one by a transformation
with the group element. In any case, the quantum theory for the open string
propagating in the SL(2,R) curved spacetime reduces to the same mathematics
outlined above, with no further consideration of left/right states since
there is a single current. So the mechanism of squaring two minus signs
hoped for through eq.(\ref{modular}) cannot help. Therefore, this model is
not unitary. Something is wrong with the open as well as the closed string!
The problem will be solved in this paper.

On the other hand one may analyze the classical theory underlying such
models in the form of a WZW or gauged WZW theory based on a non-compact
group, such that there is a single time coordinate \cite{ibcurved}. The
classical solutions in a class of models were outlined some time ago \cite
{ibsfanomaly}, and a more detailed discussion of the 2D special case has
been given more recently \cite{ibjs}\cite{folds}. The same approach yields
classical string solutions for SL(2,R) as well. For a classical string to
make sense one must consider only those solutions for which the string time
coordinate $X^0(\tau ,\sigma )$ (defined through an appropriate
coordinatization of the group element) monotonically increases as a function
of $\tau $ for the whole string (all $\sigma ),$ just as in flat spacetime.
Under such conditions one finds that there are perfectly well behaved
physical string solutions that are the generalizations of string motions in
flat spacetime\footnote{%
These solutions also seem to display new {\it classical string physics} at
singularities, such as penetration to the other side of the black hole
spacetime, which is not possible for particle geodesics (see e.g. \cite{ibjs}%
\cite{folds}).}. There must be a corresponding quantum theory formulated in
terms of the symmetry currents of the WZW theory. Evidently, it cannot be
the theory outlined above. A radical solution is needed.

\section{Modified currents}

In this paper a key departure from the usual currents is introduced in the
form of a logarithmic cut $\ln z$ in the complex $z$ plane. The new currents
satisfy the standard local operator products, with only poles as
singularities, so that the standard commutation rules are not altered. To
compensate for the cut, the physical states are restricted by monodromy
conditions.

First we explain in general terms why a cut is possible. The string
equations of motion that follow from an action (e.g. WZW or gauged WZW)
consist of differential equations and boundary conditions. The differential
equations for the left/right currents
\begin{equation}
\partial _{\bar{z}}J_L=0,\quad \partial _zJ_R=0
\end{equation}
are required to be satisfied when $z,\bar{z}$ are on the circle, i.e. $%
z=\exp (i(\tau +\sigma ))$ , $\bar{z}=\exp (i(\tau -\sigma )),\,\,$in the
Minkowski world sheet, and therefore the cut starting at $z=0,\infty $
presents no problem with the physical equations of motion. Thus the left
moving currents $J_L$ are functions of $z,$ including the possibility of $%
\ln z$ (and similarly right moving currents $J_R$ are functions of $\bar{z}).
$ Another aspect of minimizing the string action is the boundary conditions
that require periodicity in the $\sigma $ variable. The physics should be
consistent with the periodicity condition $\sigma \rightarrow \sigma +2\pi n$
$($or $z\rightarrow ze^{i2\pi n}$ in the complex $z$ plane)$.$ This is the
reason that the currents are usually taken as functions of only the powers $%
z^n$ and hence holomorphic for complex $z$, except for poles at $z=0,\infty $
(as in eq.(\ref{laurent}))$.$

The periodicity requirement will be satisfied in our solution in a more
subtle way. A cut in the currents presents problems with the monodromy.
Instead of taking periodic currents, the periodicity condition will be
implemented on the Hilbert space in our theory. So, the physical states will
be identified as the subset of states that are invariant under the
monodromy. In other words, the matrix elements of the currents will be
periodic in the physical subspace. Furthermore, we will see that although
the new currents have a logarithmic cut, the stress tensor is free of cuts
and can be written as a Laurent series in powers of $z$. This feature
permits the simultaneous imposition of the Virasoro constraints as well as
the monodromy on the states.

Except for a new definition of the currents we keep the entire formalism the
same. That is, the stress tensor will be constructed as the Sugawara form
with the new currents, and the Virasoro condition will be implemented with
this stress tensor. The structure of the new currents was discovered by
trial and error, in the process of building a new free boson realization for
SL(2,R) currents, which will be presented below. But later it was understood
that the logarithmic structure follows naturally from the WZW model. It
turns out that the natural variables for the quantization of the model, and
for performing computations, are the free bosons, rather than the currents.
However, it is instructive to state the resolution of the unitarity problem
directly in terms of currents without considering the details of free boson
realizations.

Thus define the following new set of currents constructed from the old ones
(in light-cone type combinations)
\begin{eqnarray}
\begin{array}{l}
J^0(z)+J^1(z)=\left[ \tilde J^0(z)+\tilde J^1(z)\right] \\
J^0(z)-J^1(z)=\left[ \tilde J^0(z)-\tilde J^1(z)\right] -2i\alpha _0^{-}\ln
z\,\,\tilde J^2(z) \\
\quad \quad -\frac kz\alpha _0^{-}+\left( -i\alpha _0^{-}\ln z\right)
^2\left[ \tilde J^0(z)+\tilde J^1(z)\right] \\
J^2(z)=\tilde J^2(z)-i\alpha _0^{-}\ln z\,\left[ \tilde J^0(z)+\tilde J%
^1(z)\right]
\end{array}
\label{newold}
\end{eqnarray}
The old currents $\tilde J^i(z)$ are analytic, they are written in the form
of a Laurent series as in (\ref{laurent}), and the coefficients $J_n^i$
satisfy the commutation rules in (\ref{commutators}). In addition to these
coefficients we have introduced a new zero mode $\alpha _0^{-}$ which
commutes with all the current modes $J_n^i.$ In the limit $\alpha
_0^{-}\rightarrow 0$ we get the old currents. One can compute the local
operator products and/or the commutators and show that for any $\alpha
_0^{-} $ the {\it local} commutation rules are the same for both sets of
currents
\begin{eqnarray}
\left[ \tilde J^i(z),\tilde J^j(w)\right] &=&i\epsilon ^{ijk}\eta _{kl}%
\tilde J^l(w)\,\delta (z-w)  \nonumber \\
&&-\frac k2\partial _z\delta (z-w)\eta ^{ij}, \\
\left[ J^i(z),J^j(w)\right] &=&i\epsilon ^{ijk}\eta _{kl}J^l(w)\,\delta (z-w)
\nonumber \\
&&-\frac k2\partial _z\delta (z-w)\eta ^{ij}.  \nonumber
\end{eqnarray}
The first line is given, and is equivalent to the commutation rules of the
current modes $J_n^i$ given in (\ref{commutators}). The second line is
derived from the first line plus the definition of the new currents. The $%
\ln z$ structure plays a non-trivial role in arriving at this result. In
particular the crucial commutator to check is
\begin{equation}
\left[ J^2(z),\,\,J^0(w)-J^1(w)\right] =-i\delta (z-w)\,\left(
J^0(w)-J^1(w)\right)
\end{equation}
where the term $-i$ $\delta (z-w)\,\left( -\frac kw\alpha _0^{-}\right) $ on
the right hand side must be reproduced as part of the new currents. This
term is obtained from the combination of the central extensions coming from
\begin{eqnarray}
&&\ \ \left[ \tilde J^2(z),-2i\alpha _0^{-}\ln w\,\,\tilde J^2(w)\right]
\quad \quad \rm{and} \\
&&\ \ \left[ -i\alpha _0^{-}\ln z\,\left( \tilde J^0(z)+\tilde J^1(z)\right)
,\,\left( \tilde J^0(z)-\tilde J^1(z)\right) \right]  \nonumber
\end{eqnarray}
Collecting the central extensions of these two terms we have
\begin{eqnarray}
&&-\frac k2\partial _z\delta (z-w)\,\left\{ -2i\alpha _0^{-}\ln w+2i\alpha
_0^{-}\ln z\right\}  \nonumber \\
&=&\frac{ik\alpha _0^{-}}w\delta (z-w)
\end{eqnarray}
which gives the desired result. Thus, even though the new currents have
different global properties on the world sheet, they have the same local
singularities as the old currents when their products are considered. The
global structure has an effect on the spectrum of the theory through the
monodromy, as will be discussed in section VIII.

We now claim that the physical model has the new currents as the symmetry
currents, and that the stress tensor is constructed from the new currents
\begin{equation}
T(z)=\frac 1{k-2}:\left( -\left( J_0(z)\right) ^2+\left( J_1(z)\right)
^2+\left( J_2(z)\right) ^2\right) :
\end{equation}
where normal ordering is defined by splitting the points, subracting the
singularity, and then sending $z\rightarrow w$ in the finite part. This
procedure produces the following expression when written in terms of the old
currents
\begin{equation}
\begin{array}{ll}
T(z)= & \frac 1{k-2}:\left( -\left( \tilde J_0(z)\right) ^2+\left( \tilde J%
_1(z)\right) ^2+\left( \tilde J_2(z)\right) ^2\right) : \\
& \quad +\frac 1z\alpha _0^{-}\,\left( \tilde J^0(z)+\tilde J^1(z)\right)
\end{array}
\end{equation}
This implies that the Virasoro constraints are modified as follows\footnote{%
This form of stress tensor was considered before from a different point of
view \cite{halpern}, but apparently without realizing that it comes from a
current with a logarithmic singularity, and that it implies monodromy
constraints on the states.}
\begin{equation}
L_n=\tilde L_n+\alpha _0^{-}\,\left( J_n^0+J_n^1\right) .  \label{newvir}
\end{equation}
Using the algebra (\ref{commutators}) for $\tilde L_n,J_n^i$ it is easily
shown that the central charge of the new $L_n$ is the one given before (for
any $\alpha _0^{-})$
\begin{equation}
\left[ L_n,L_m\right] =(n-m)L_{n+m}+\frac c{12}n(n^2-1)\delta _{n+m}.
\end{equation}
The new structure (\ref{newvir}) changes the physical conditions on the
Hilbert space in a profound way. Furthermore, instead of using the basis $%
|jm>$ in which the zero mode $J_0^0$ is diagonal, it is necessary to use the
basis $|j,p^{+},p^{-}>$ in which $J_0^0+J_0^1=p^{+}$ and $\alpha
_0^{-}=p^{-} $ are diagonal in order to diagonalize $L_0.$ The eigenvalue of
the Casimir is insensitive on which zero mode generator is diagonalized.
Then the eigenvalue of $L_0$ is
\begin{equation}
L_0=p^{-}p^{+}-\frac{j(j+1)}{k-2}+l=a  \label{balance}
\end{equation}
with $a\leq 1.$ The contribution from $\alpha _0^{-}=p^{-}$ alters the
balance of this equation as compared to (\ref{balanc}). When $p^{-}$ was
absent $j(j+1)$ had to be positive, which in turn required the discrete
series. However, now $p^{-}p^{+}=-M^2/2$ plays the role of a mass$^2$ in a
2-dimensional subspace, and therefore $j(j+1)$ can be negative . In that
case the principal as well as the supplementary series become relevant in
the description of excited string states. This is basically the way out of
the bind with the ghosts. The computations that led to the negative norm
states were valid only for the discrete series, but now we have other
choices of unitary representations of SL(2,R) at the base. Furthermore, the
new Virasoro constraints require physical states that correspond to a
different combination of Verma module states. One may compute as in the past
``physical'' states and then check their norms. However, this procedure is
very cumbersome. Furthermore, it may not be very useful to consider the
representation space of the old current algebra, since from that point of
view the states do not fall into degenerate representations of the central
SL(2,R) generated by $J_0^i$. The stress tensor is covariant under the
transformations generated by the new currents, and the old symmetry looks
like spontaneously broken (as in (\ref{newvir})). Therefore, in order to
solve the new theory we must resort to better methods. This will be done
below, where we will show that the theory can be solved completely in terms
of free fields. In particular we will show that in the free field
representation that corresponds to the WZW model only the principal series
is selected, and that there are no ghosts after the Virasoro constraints are
satisfied.

\section{SL(2,R) Currents and free fields}

\subsection{Definition of the fields}

Consider the free fields $X^{-}(z),\,P^{+}(z),$ $\,S(z),\,T^{\prime }(z)$.
They have na\"\i ve dimensions 0,1,1,2 respectively and they are defined as
follows

\begin{eqnarray}
X^{-}(z) &=&q^{-}-i\alpha _0^{-}\,\ln z+i\sum_{n\neq 0}\frac 1n\alpha
_n^{-}\,z^{-n},\,\quad \left( \alpha _n^{-}\right) ^{\dagger }=\alpha
_{-n}^{-}  \\
P^{+}(z) &=&\sum_{n=-\infty }^\infty \alpha _n^{+}\,z^{-n-1},\,\quad \left(
\alpha _n^{+}\right) ^{\dagger }=\alpha _{-n}^{+}  \nonumber \\
S(z) &=&\sum_{n=-\infty }^\infty s_n\,z^{-n-1},\quad \left( s_n\right)
^{\dagger }=s_{-n}  \label{fields} \\
T^{\prime }(z) &=&\sum_{n=-\infty }^\infty L_n^{\prime }\,z^{-n-2},\quad
\left( L_n^{\prime }\right) ^{\dagger }=L_{-n}^{\prime }  \nonumber
\end{eqnarray}
These fields are Hermitean (defined for any complex $z$ by)
\[
\left( X^{-}(z)\right) ^{\dagger }=X^{-}(\frac 1z),\quad \left(
zP^{+}(z)\right) ^{\dagger }=\frac 1zP^{+}(\frac 1z),\quad etc.
\]
That is, due to the hermiticity of the modes, the fields $%
X^{-}(z),\,z\,P^{+}(z),$ $\,zS(z),\,z^2T^{\prime }(z)$ are Hermitian if $z$
is on the unit circle, as it would be for describing left/right moving
fields $z=\exp (i(\tau \pm \sigma ))$ in the Minkowski world sheet. We
emphasize that under Hermitean conjugation the $\alpha _n^{+}$ modes do not
go to the $\alpha _n^{-}$ modes. The zero modes $q^{-},\alpha _0^{\pm
}=p^{\pm },$\thinspace $s_0,\,L_0^{\prime }$ play an important role, as we
will see below. We also assign the commutation rules
\begin{equation}
\begin{array}{l}
\lbrack q^{-},\alpha _0^{+}]=i\,, \\
\left[ \alpha _n^{-}\,,\alpha _m^{+}\right] =n\,\delta _{n+m,0}\,\,\,, \\
\left[ s_n\,,s_m\right] =\left( \frac k2-1\right) \,n\,\delta _{n+m,0}\,\,\,,
\\
\left[ L_n^{\prime }\,,L_m^{\prime }\right] =(n-m)L_{n+m}^{\prime }
\end{array}
\label{commuta}
\end{equation}
while all other commutators are zero. In particular all $\alpha _n^{+}$
commute among themselves, and all $\alpha _n^{-}$ commute among themselves,
just like lightcone coordinates. Indeed the $\alpha _n^{\pm }$ oscillators
may be rewritten in terms of light-cone type combinations of one time-like $%
\alpha _n^0$ and one space-like $\alpha _n^1$ oscillator, i.e. $\alpha
_n^{\pm }=(\alpha _n^1\pm \alpha _n^0)/\sqrt{2}.$ In this sense, $%
q^{-},p^{\pm }$ are interpreted as light-cone type canonical variables $%
q=x^{-},\,p=p^{+}.$ We have {\it not} introduced a canonical variable
corresponding to $x^{+},$ hence $\alpha _0^{-}=p^{-}$ commutes with all$\,$
the operators and acts like a constant. Similarly the zero mode $s_0$ also
acts like a constant.

The $L_n^{\prime }$ operators act like Virasoro operators with zero central
charge. The minimal construction of SL(2,R) currents does not need these
operators. In fact, the WZW model that we discuss in section VII does not
have the $L_n^{\prime }$. But in this section we would like to to present a
new more general construction of SL(2,R) currents that include the $%
L_n^{\prime }$ since they may find physical applications in more general
models. In their absence only the principal series arises, but in their
presence all unitary representations of SL(2,R) current algebra can be
constructed, as will be seen below.

The field $X^{-}(z)$ contains the logarithmic term $-i\alpha _0^{-}\ln z.\,$
This is the source of the $\ln z$ terms in the currents discussed in the
previous section. If this term is left out all the expressions are analytic,
and we obtain the old currents. However, the presence of this term is
crucial for the description of excited strings in a unitary theory.

Normal ordering is defined in the usual way for all the oscillators, i.e.
positive modes are moved to the right and negative modes to the left.
However, for the zero modes $q^{-},p^{+}$ we define normal ordering to mean
the Hermitean combination of any powers of $q^{-},p^{+}$, for example
\begin{eqnarray}
\ &:&q^{-}p^{+}:\,\equiv \frac 12(q^{-}p^{+}+p^{+}q^{-}), \\
\ &:&q^{-}q^{-}p^{+}:\,\equiv q^{-}p^{+}q^{-},  \label{normalherm} \\
\ &:&q^{-}p^{+}q^{-}p^{+}:\,\equiv \frac 12\left(
q^{-}p^{+}q^{-}p^{+}+p^{+}q^{-}p^{+}q^{-}\right) ,\,\,\,etc.  \nonumber
\end{eqnarray}
This is important for constructing Hermitean currents. We define the
contractions $<q^{-}p^{+}>=i/2$ and $<p^{+}q^{-}>=-i/2$ which arise in
rewriting ordinary products in terms of Hermitean products
\begin{eqnarray}
q^{-}p^{+} &=&\,:q^{-}p^{+}:\,+\,<q^{-}p^{+}>,\,\quad <q^{-}p^{+}>=i/2, \\
p^{+}q^{-} &=&\,:q^{-}p^{+}:\,+\,<p^{+}q^{-}>,\,\quad <p^{+}q^{-}>=-i/2.
\nonumber
\end{eqnarray}
With this definition of normal ordering the usual Wick's theorem for
multiplying normal ordered products with each other is preserved. Then, one
obtains, for example
\begin{equation}
\begin{array}{l}
:q^{-}p^{+}:\times :q^{-}p^{+}:\,=\,:q^{-}p^{+}q^{-}p^{+}: \\
\quad \quad +\,<p^{+}q^{-}><q^{-}p^{+}> \\
\quad \quad +\left( <p^{+}q^{-}>+<q^{-}p^{+}>\right) \,:q^{-}p^{+}: \\
\quad =\,:q^{-}p^{+}q^{-}p^{+}:\,+\frac 14.
\end{array}
\end{equation}

Products of fields may be rewritten in the normal ordered form as follows
\begin{equation}
A(z)\,B(w)=\,:A(z)\,B(w):+<A(z)\,B(w)>\,\,,  \label{prod}
\end{equation}
The following $c-terms$ arise from the normal ordering of the various fields
\begin{eqnarray}
\ &<&X^{-}(z)\,P^{+}(w)>\,\equiv \frac i{2w}+\frac i{z-w}  \nonumber \\
\ &<&P^{+}(z)X^{-}(w)\,>\,\equiv \frac i{2z}+\frac{-i}{z-w}
\label{contractions} \\
\ &<&S(z)\,S(w)>\,\equiv \frac{k/2-1}{\left( z-w\right) ^2}  \nonumber
\end{eqnarray}
The $\frac i{2w}\,$or $\frac i{2z}$ terms occur because of the unusual
normal ordering of the zero modes $p^{+},q^{-}.$ These are unimportant in
the computation of the singular parts of operator products, but they do play
a role in the computation of the finite parts, such as the energy-momentum
tensor, as seen below. The presence or absence of the $p^{-}\ln z$ term in
the definition of $X^{-}(z)$ does not change the contractions in (\ref
{contractions}), therefore the singularity structure in the operator
products are unaffected by the $p^{-}\ln z$ term. However, the presence of
the $p^{-}$ term does change the finite parts in a desirable way.

\subsection{Construction of the currents}

Some time ago we suggested a free field construction of SL(2,R) currents
that gives only the principal series at the base \cite{ibstonybrook}. This
was similar to the Wakimoto construction but with the important difference
that the currents were Hermitean. Here we generalize that construction by
including all unitary representations of SL(2,R) at the base. In this
construction there are new structures that have not been introduced
heretofore. After a few trials and errors we found that the following works
\begin{eqnarray}
J_0(z)+J_1(z) &=&\,P^{+}(z) \\
J_0(z)-J_1(z) &=&\,:X^{-}(z)\,P^{+}(z)\,X^{-}(z):+2S(z)\,X^{-}(z)\nonumber \\
&&-\,\,ik\partial _zX^{-}(z)-\frac{(k-2)T^{\prime }(z)}{P^{+}(z)}
\label{currents} \\
J_2(z) &=&\,:X^{-}(z)\,P^{+}(z):+S(z)  \nonumber
\end{eqnarray}
All the currents are Hermitean.
\begin{equation}
\left( zJ^\mu (z)\right) ^{\dagger }=\frac 1zJ^\mu (\frac 1z)\,.
\label{hermitcurrent}
\end{equation}
The new aspects include the $\ln z$ terms and the $T^{\prime }(z)/P^{+}(z)$
parts. $T^{\prime }$ is a stress tensor with {\it zero central charge. }It
commutes with all other terms in these currents{\it , }but it obeys the
following operator product with itself
\begin{equation}
T^{\prime }(z)\times T^{\prime }(w)=\frac 0{\left( z-w\right) ^4}+\frac{%
2T^{\prime }(w)}{\left( z-w\right) ^2}+\frac{\partial _wT^{\prime }(w)}{z-w}%
+\cdots
\end{equation}
It is possible to give a construction of $T^{\prime }(z)$ (or $L_n^{\prime }$%
) in terms of other elementary (free) fields, but this is not necessary for
the present paper since it will be used only in the form of $T^{\prime }(z)$.

Due to the $p^{-}\ln z$ term in $X^{-}(z)$, the currents can not be written
purely as a Laurent series. Let us define another set of currents $\tilde J%
^i $ as the $\alpha _0^{-}=0$ limit of the currents above. The $\tilde J%
^i(z) $ have the Laurent expansion with Hermitean coefficients
\begin{equation}
\tilde J^i(z)=\sum J_n^i\,z^{-n-1},\quad \left( J_n^i\right) ^{\dagger
}=J_{-n}^i.
\end{equation}
Then the $J_n^i$ become functions of the oscillators $\alpha _n^{\pm },s_n$
and $L_n^{\prime }$. The relation between the new non-analytic currents $%
J^i(z)$ and the analytic currents $\tilde J^i(z)$ has precisely the form
given in eq.(\ref{newold}).

Using the rules for normal ordering given above we compute the operator
product algebra for the currents, and verify that they satisfy the correct
relations in the presence or absence of the $\alpha _0^{-}\ln z$ term
\begin{eqnarray}
\tilde J^\mu (z)\,\tilde J^\nu (w) &\rightarrow &\frac{k/2}{\left(
z-w\right) ^2}+i\epsilon ^{\mu \nu \lambda }\,\frac{\tilde J_\lambda \left(
w\right) }{z-w}+\cdots  \label{covariantprod} \\
J^\mu (z)\,J^\nu (w) &\rightarrow &\frac{k/2}{\left( z-w\right) ^2}%
+i\epsilon ^{\mu \nu \lambda }\,\frac{J_\lambda \left( w\right) }{z-w}+\cdots
\end{eqnarray}
(see section V and the Appendix for the details of the calculation).

The energy momentum tensor is obtained from the normal ordered product of
the currents
\begin{equation}
T(z)=\frac 1{k-2}:\left( -\left( J_0(z)\right) ^2+\left( J_1(z)\right)
^2+\left( J_2(z)\right) ^2\right) :  \label{stresst}
\end{equation}
The result of the computation gives (see section VI)
\begin{equation}
T(z)=\,:P^{+}i\partial _zX^{-}:+T_S(z)+T^{\prime }(z)\quad ,
\label{stresstensor}
\end{equation}
If the computation is repeated with the $\tilde J(z)$ currents the only
difference is dropping the $\alpha _0^{-}$ term contained in
\begin{equation}
i\partial _zX^{-}=\sum_{n=-\infty }^\infty \alpha _n^{-}z^{-n-1}.
\end{equation}
In (\ref{stresstensor}) $T_S$ is a Hermitean stress tensor
\begin{equation}
T_S(z)=\frac 1{k-2}\left[ :\left( S(z)\right) ^2:-\,\,\frac iz\partial
_z\left( zS(z)\right) +\frac 1{4z^2}\right]  \label{ts}
\end{equation}
The structure $\frac iz\partial _z\left( zS(z)\right) $ differs from the
usual one $i\partial S,$ and thus is Hermitean. The operator products of $%
T_S(z)$ are

\begin{equation}
T_S(z)\times T_S(w)=\frac{c_s/2}{\left( z-w\right) ^2}+\frac{2T_S(w)}{\left(
z-w\right) ^2}+\frac{\partial _wT_S(w)}{z-w}+\cdots
\end{equation}
with the central charge
\begin{equation}
c_s=1+\frac 6{k-2}.
\end{equation}
Note that the term $P^{+}i\partial _zX^{-}$ is identical to the energy
momentum tensor of flat light-cone coordinates constructed from the
oscillators $\alpha _n^{\pm }$. Therefore, that part is mathematically
equivalent to a $c=2$ stress tensor constructed from one time and one space
coordinate in flat spacetime. Then the total central charge is
\begin{eqnarray}
c &=&2+c_s+c^{\prime }  \nonumber \\
\ &=&2+\left( 1+\frac 6{k-2}\right) +0 \\
\ &=&\frac{3k}{k-2},  \nonumber
\end{eqnarray}
which is the right central charge for the $SL(2,R)$ WZW model. Finally, as a
further consistency check, by using only the operator products of the
elementary fields, one finds that $T(z)$ has the correct operator products
with the currents.

The zero mode of the stress tensor takes the form $L_0=L_0^{\pm
}+L_0^S+L_0^{\prime },$ where each piece has the eigenvalues
\begin{eqnarray}
L_0^{\pm } &=&p^{+}p^{-}+l_{\pm },  \nonumber \\
L_0^S &=&(s_0^2+1/4)/(k-2)+l_s, \\
L_0^{\prime } &=&h^{\prime }+l^{\prime }.  \nonumber
\end{eqnarray}
where $l_{\pm },l_s,l^{\prime }$ are positive integers and $h^{\prime }$ is
the eigenvalue of $L_0^{\prime }$ at the base (whose possible values depend
on the model for $T^{\prime })$. The mass shell condition $L_0=a$ is
\begin{equation}
p^{+}p^{-}+(s_0^2+1/4)/(k-2)+h^{\prime }+%
\mathop{\rm integer}
=a  \label{balancee}
\end{equation}
where $a\leq 1.$ The term $p^{+}p^{-}=-M^2/2$ is crucial since it takes
negative values. In comparison to eq.(\ref{balance}) we see that the Casimir
of the old currents takes the value
\begin{equation}
j(j+1)=-(s_0^2+1/4)-h^{\prime }(k-2).
\end{equation}
Therefore, if the $T^{\prime }$ piece is absent in the construction ($%
h^{\prime }=0),$ then $j=-1/2\pm is_0$ is only in the principal series. The
supplementary series occurs for -$-1/4<j(j+1)<0$ and the discrete series
occurs for $-1/4<j(j+1).\,$ We see that the field $T^{\prime }$ with a
positive $h^{\prime }$ contributes only to the principal series and with a
negative $h^{\prime }$ it leads to the other representations as well. This
construction may find various applications in the future. We will see below
that $T^{\prime }$ is absent in the SL(2,R) WZW model, hence only the
special case of our construction ($T^{\prime }=0)\,$ will find an
application in the current paper. Then, for excited string states, since the
integer in (\ref{balancee}) is positive it would not be possible to satisfy
the mass shell condition in the absence of the $p^{-}.$ So, the new
logarithmic structure will play a role.

The physical state conditions will be analyzed in section VIII, after we
prove the above construction.

\section{Operator products of currents}

We want to verify that the operator products of the currents are correct.
Some formulas that are useful for this computation are collected in the
appendix. There are 6 independent operator products that we need to
reproduce. In increasing complexity these are
\begin{equation}
\begin{array}{l}
\left( J_0+J_1\right) (z)\times \left( J_0+J_1\right) (w)\rightarrow 0+\cdots
\\
\left( J_0+J_1\right) (z)\times J_2(w)\rightarrow \frac{-i}{z-w}\left(
J_0+J_1\right) (w)+\cdots \\
\left( J_0+J_1\right) (z)\times \left( J_0-J_1\right) (w)\rightarrow \frac{-k%
}{\left( z-w\right) ^2}-\frac{2i}{z-w}J_2(w)+\cdots \\
J_2(z)\times J_2(w)\rightarrow \frac{k/2}{\left( z-w\right) ^2}+\cdots \\
J_2(z)\times \left( J_0-J_1\right) (w)\rightarrow \frac{-i}{z-w}\left(
J_0-J_1\right) (w)+\cdots \\
\left( J_0-J_1\right) (z)\times \left( J_0-J_1\right) (w)\rightarrow 0+\cdots
\end{array}
\label{explprod}
\end{equation}
The fist one is easy
\begin{eqnarray}
&&\left( J_0(z)+J_1(z)\right) \times \left( J_0(z)+J_1(z)\right)
\label{j01j01} \\
&=&P^{+}(z)\times P^{+}(w)\rightarrow \,0  \nonumber
\end{eqnarray}
This is correct since the central extension in $J_0\times J_0$ cancels the
central extension in $J_1\times J_1$ due to the indefinite metric $\eta
_{\mu \nu }=diag(-1,1,1).$ The product
\begin{eqnarray}
\left( J_0+J_1\right) (z)\times J_2(w) &=&P^{+}(z)\times
[:X^{-}\,P^{+}:+S](w)  \\
&\rightarrow &-iP^{+}(w)/\left( z-w\right)  \label{j01j2} \\
&\rightarrow &-i\left( J_0(w)+J_1\left( w\right) \right) /(z-w)  \nonumber
\end{eqnarray}
follows from eq.(\ref{pxqp}). Next, by using the operator product for $P^{+}$
and $X^{-}$ one obtains
\begin{equation}
\begin{array}{l}
\left( J_0+J_1\right) (z)\times \left( J_0-J_1\right) (w) \\
=P^{+}(z)\times \left[
\begin{array}{c}
:X^{-}\,P^{+}X^{-}:-ik\partial _wX^{-} \\
+2X^{-}S-%
{\textstyle {(k-2)T^{\prime } \over P^{+}}}
\end{array}
\right] (w) \\
\rightarrow \frac{-k}{\left( z-w\right) ^2}+\frac{-2i}{z-w}%
[:X^{-}\,P^{+}:+S](w) \\
\rightarrow \frac{-k}{\left( z-w\right) ^2}+\frac{-2i}{z-w}J_2(w).
\end{array}
\label{j01j0m1}
\end{equation}
The product for $J_2\times J_2$ is obtained by using (\ref{qpxqp}) and (\ref
{contractions})
\begin{equation}
\begin{array}{l}
J_2(z)\times J_2(w) \\
=[:X^{-}\,P^{+}:+S](z)\times \left[ :X^{-}\,P^{+}:+S\right] (w) \\
\rightarrow \frac 1{\left( z-w\right) ^2}+\frac{k/2-1}{\left( z-w\right) ^2}=%
\frac{k/2}{\left( z-w\right) ^2}.
\end{array}
\label{j2j2}
\end{equation}
The product $J_2\times \left( J_0-J_1\right) $ is computed by using (\ref
{qpxqpq}) and (\ref{contractions})
\begin{equation}
\begin{array}{l}
J_2(z)\times \left( J_0-J_1\right) (w)= \\
_{=\,\left( :X^{-}\,P^{+}:+S\right) (z)\times \,\left[
:X^{-}P^{+}X^{-}:\,-ik\partial _wX^{-}+2X^{-}S-%
{\textstyle {(k-2)T^{\prime } \over P^{+}}}
\right] (w)} \\
\rightarrow \frac{-i\,\left( :X^{-}P^{+}X^{-}-(k-2)T^{\prime }/P^{+}\right)
(w)}{z-w}+\frac{-2iX^{-}(w)}{z-w}S(w) \\
\quad +\frac{2X^{-}(w)}{\left( z-w\right) ^2}-ikX^{-}(z)\left( \partial _w%
\frac{-i}{z-w}\right) +2X^{-}(w)\frac{k/2-1}{\left( z-w\right) ^2} \\
\rightarrow \frac{-i}{z-w}\left[ :X^{-}\,P^{+}X^{-}:-ik\partial
_wX^{-}+2X^{-}S-%
{\textstyle {(k-2)T^{\prime } \over P^{+}}}
\right] \\
\rightarrow \frac{-i}{z-w}\left( J_0-J_1\right) (w)
\end{array}
\label{j2j0m1}
\end{equation}
where we have used
\begin{equation}
X^{-}(z)=X^{-}(w)+\left( z-w\right) \partial _wX^{-}(w)+\cdots  \label{logs}
\end{equation}
in the fourth line. Finally the product that is the most complicated to
compute follows from the steps that lead to (\ref{j01xj01})
\begin{equation}
\left( J_0-J_1\right) (z)\times \left( J_0-J_1\right) (w)\rightarrow 0.
\end{equation}
Hence we have correctly constructed the current algebra from the elementary
free fields. The closure of the algebra goes through whether or not the $%
\alpha _0^{-}\ln z$ term is included in the definition of $X^{-}(z).$ Its
presence is felt through eq.(\ref{logs})\thinspace since this is how the $%
1/z $ terms work out. Thus, for any $\alpha _0^{-}$ the algebra of the new
currents is identical to the algebra of the old ones.

\section{The stress tensor}

To construct the stress tensor we need to compute the normal ordered
products of currents $\eta _{ij}:J^i(z)J^j(z):$. We define the normal
ordered product by splitting the points, and taking the ordinary product
minus the singularity. That is,
\begin{equation}
\eta _{ij}:J^i(z)J^j(w):\,\equiv \eta _{ij}\,J^i(z)J^j(w)-\frac{3k/2}{\left(
z-w\right) ^2}.  \label{prodmsing}
\end{equation}
Then we compute the right hand side by substituting the free field form and
rearrange it by using Wick's theorem for free fields. The final form is a
normal ordered form for the free fields in which all singularities cancel.
Then the limit $z\rightarrow w$ is taken to define the local stress tensor.
In this process it is important to keep track of the finite parts, and not
use the operator products na\"\i vely. We need to compute the products $%
\frac 12\left( J_0-J_1\right) (z)\times \left( J_0+J_1\right) (w)$ , $\frac 1%
2\left( J_0+J_1\right) (z)\times \left( J_0-J_1\right) (w)$ , and $%
J_2(z)\times J_2(w)$ including the finite parts. The singular part for the
first two products is given in (\ref{j01j0m1}) and the finite part (as $%
z\rightarrow w)$ is obtained from eqs.(\ref{contractions},\ref{pxqpq}).
Thus,
\begin{equation}
\begin{array}{l}
-\frac 12\left( J_0+J_1\right) (z)\times \left( J_0-J_1\right) (w) \\
_{=-\frac 12P^{+}(z)\times \,\left[ :X^{-}\,P^{+}X^{-}:\,-ik\partial
_wX^{-}+2X^{-}S-(k-2)\frac{T^{\prime }}{P^{+}}\right] (w)} \\
_{=-\frac 12\left( :P^{+}X^{-}P^{+}X^{-}:\right) +\frac{ik}2\left(
:P^{+}\partial _wX^{-}:\right) -:P^{+}X^{-}:S+(\frac k2-1)T^{\prime }} \\
\quad +\frac k{2\left( z-w\right) ^2}-\left( \frac{-i}{z-w}+\frac i{2w}%
\right) J_2(w)+O(z-w)
\end{array}
\end{equation}
Here we were careful to keep the finite $1/z$ part in the contraction $%
<P^{+}(z)X^{-}(w)>.$ Similarly we have
\begin{equation}
\begin{array}{l}
-\frac 12\left( J_0-J_1\right) (z)\times \left( J_0+J_1\right) (w) \\
_{=-\frac 12\left[ :X^{-}\,P^{+}X^{-}:\,-ik\partial _zX^{-}+2X^{-}S-(k-2)%
\frac{T^{\prime }}{P^{+}}\right] (z)\,\times P^{+}(w)} \\
_{=-\frac 12\left( :X^{-}P^{+}X^{-}P^{+}:\right) +\frac{ik}2\left(
:P^{+}\partial _wX^{-}:\right) -:P^{+}X^{-}:S+(\frac k2-1)T^{\prime }} \\
_{\quad +}\frac k{2\left( z-w\right) ^2}-\left( \frac i{z-w}+\frac i{2w}%
\right) J_2(w)-i\partial _wJ_2(w) \\
\quad \quad +O(z-w)
\end{array}
\end{equation}
The singular part of the product $J_2(z)\times J_2(w)$ is given in (\ref
{j2j2}) while the finite part is obtained through eqs.(\ref{qpxqp}) and (\ref
{contractions}) as follows
\begin{equation}
\begin{array}{l}
J_2(z)\times J_2(w)= \\
=\left( :X^{-}\,P^{+}:\,+S\right) (z)\times \left( :X^{-}\,P^{+}:\,+S\right)
(w) \\
=\,:X^{-}(z)\,P^{+}(z)X^{-}(w)\,P^{+}(w): \\
\quad +\left( \frac{-i}{z-w}+\frac i{2z}\right) :X^{-}(z)\,P^{+}(w): \\
\quad +\left( \frac i{z-w}+\frac i{2w}\right)
:X^{-}(w)\,P^{+}(z):+2S:X^{-}P^{+}: \\
\quad +\left( \frac i{z-w}+\frac i{2w}\right) \left( \frac{-i}{z-w}+\frac i{%
2z}\right) +:S(z)S(w):+\frac{k/2-1}{\left( z-w\right) ^2} \\
\rightarrow \frac k{2\left( z-w\right) ^2}-\frac 1{4w^2}+\,\left( :i\partial
P^{+}X^{-}-P^{+}i\partial X^{-}:\right) \\
\quad +\frac iw:X^{-}P^{+}:+:X^{-}\,P^{+}X^{-}\,P^{+}: \\
\quad \quad +:SS:+2S:X^{-}P^{+}:+O(z-w)
\end{array}
\end{equation}
where we have used
\begin{equation}
-\frac 1{2z}\frac 1{z-w}=-\frac 1{2w}\frac 1{z-w}+\frac 1{2w^2}+O(z-w)
\end{equation}
in arriving at the $1/4w^2$ term. Substituting these expressions in (\ref
{prodmsing}) we obtain, as $z\rightarrow w$
\begin{eqnarray}
\frac{\eta _{ij}}{k-2} &:&J^i(w)J^j(w):=\left( :P^{+}i\partial
_wX^{-}:+T_S+T^{\prime }\right) \,  \nonumber \\
T_S &=&\frac 1{k-2}\left( :S^2:\,-\frac iw\partial \left( wS\right) +\frac 1{%
4w^2}\right)
\end{eqnarray}
Therefore the energy momentum tensor is
\begin{equation}
T=\,:P^{+}i\partial X^{-}:+T_S+T^{\prime }
\end{equation}
as advertized in eq.(\ref{stresstensor}).

\section{Quantization of the SL(2,R) WZW model}

Up to now we have worked in a purely algebraic framework. We now relate
these structures to the WZW model for SL(2,R). The group element $g(X)$ may
be parametrized in terms of 1-time and 2-space string coordinates $X^\mu
(\tau ,\sigma ),\,\,\mu =0,1,2.$ When the model is explicitly written in
terms of these, the action looks like eq.(\ref{action}) with the metric $%
G_{\mu \nu }(X)$ induced by the Cartan connection of the group as follows
\begin{eqnarray}
dg(X)\,g^{-1}(X) &=&\left( -it_i\right) \,dX^\mu \,E_\mu ^i(X)\, \\
G_{\mu \nu }(X) &=&E_\mu ^i(X)\,\eta _{ij}\,E_\nu ^j(X),  \nonumber
\end{eqnarray}
where $t_i$ is a basis for the SL(2,R) Lie algebra, and $\eta
_{ij}=-2tr(t_it_j)=diag(-1,1,1).$ A convenient basis that we will use is
given in terms of Pauli matrices $t_0=\sigma _2/2,\,\,t_1=i\sigma
_1/2,\,\,t_2=-i\sigma _3/2.$

The quantum theory is conveniently formulated in terms of the left and right
moving currents after writing $g(\tau ,\sigma )=g_L(\tau +\sigma
)\,g_R^{-1}(\tau -\sigma )$
\begin{equation}
J_L(z)=ik\partial _zg_Lg_L^{-1},\quad J_R(\bar z)=ik\partial _{\bar z%
}g_Rg_R^{-1},  \label{clcur}
\end{equation}
where $z=e^{i(\tau +\sigma )},\,\bar z=e^{i(\tau -\sigma )}.\,\,$Note that $%
z,\bar z$ are independent complex variables. After normal ordering, the
stress tensor takes the form
\begin{eqnarray}
T(z) &=&\frac 1{2(k-2)}tr(:J_L(z)J_L(z):), \\
\bar T(\bar z) &=&\frac 1{2(k-2)}tr(:J_R(\bar z)J_R(\bar z):).  \nonumber
\end{eqnarray}
The quantum rules are most conveniently given in terms of operator products
among the currents and the group elements. The left-movers $J_L(z),\,g_L(z)$
or right movers $J_R(\bar z),\,g_R(\bar z)$ obey similar rules. Therefore,
to save space, in the following we denote $J(z)$ and $g(z)$ for either the
left-moving or right moving pair
\begin{eqnarray}
J^i(z)\,J^j(w) &\rightarrow &\frac{k/2}{\left( z-w\right) ^2}+i\epsilon
^{ijk}\,\frac{J_k\left( w\right) }{z-w}+\cdots  \label{quantumrules} \\
J^i(z)\,g(w) &\rightarrow &\frac{-t^i}{z-w}g(w)+\cdots  \nonumber
\end{eqnarray}
where $\cdots $ stand for non-singular terms in the operator product. These
quantum rules reflect the left/right symmetry structure that is of
fundamental importance in this model. Below we give a construction in terms
of oscillators that satisfies these rules.

To obtain a relation to the free boson realization discussed in the previous
sections we coordinatize the group element (i.e. $g_L$ or $g_R$) in terms of
triangular and diagonal matrices as follows
\begin{equation}
g(z)=\left(
\begin{array}{cc}
1 & 0 \\
X^{-} & 1
\end{array}
\right) \left(
\begin{array}{cc}
:e^{-\frac u{k-2}}: & 0 \\
0 & :e^{\frac u{k-2}}:
\end{array}
\right) \left(
\begin{array}{cc}
1 & X^{+} \\
0 & 1
\end{array}
\right)  \label{group}
\end{equation}
and compute the currents
\begin{equation}
i\left( k-2\right) :\partial _zgg^{-1}:\,=\left(
\begin{array}{cc}
-J^2(z) & J^0(z)+J^1(z) \\
-J^0(z)+J^1(z) & J^2(z)
\end{array}
\right)  \label{cur}
\end{equation}
As compared to the classical currents (\ref{clcur}) we have shifted $%
k\rightarrow \left( k-2\right) \,$ in both $g$ (\ref{group}) and the
definition of the current (\ref{cur}), and applied normal ordering. This
renormalization is necessary for the commutation rules to work out, and is
consistent with similar phenomena concerning the quantization of the WZW
model \cite{ibsfeffaction}. One finds
\begin{eqnarray}
J^0(z)+J^1(z) &=&\,:\left( k-2\right) \,i\partial _zX^{+}\,e^{-2u/(k-2)}:
\nonumber \\
J^2(z) &=&\left( k-2\right) :i\partial _zX^{+}X^{-}e^{-2u/(k-2)}:+i\partial
_zu  \nonumber \\
J^0(z)-J^1(z) &=&\left( k-2\right) :i\partial _zX^{+}\left( X^{-}\right)
^2e^{-2u/(k-2)}:  \nonumber \\
&&\ \ \ \quad \quad +2X^{-}i\partial _zu-ik\partial _zX^{-}
\end{eqnarray}
The coefficient of $-ik\partial _zX^{-}$ is ambiguous because of the normal
ordering of the term $:i\partial _zX^{+}\left( X^{-}\right)
^2e^{-2u/(k-2)}:\,$. Again this has to be fixed by requiring that the
commutation rules work out. Therefore, instead of having naively $-i\left(
k-2\right) \partial _zX^{-},$ we actually must have $-ik\partial _zX^{-}.$
These results are established by applying the canonical formalism and
identifying these structures with canonical conjugate variables. Velocities
must be replaced by canonical momenta. Note that for left/right movers $%
\partial _z$ can be related to time derivatives $\partial _\tau $ or space
derivatives $\partial _\sigma $. So, at the quantum level we find that we
must identify the canonical pairs ($X^{-},P^{+})$ and $\left( u,S\right) $
as follows
\begin{eqnarray}
P^{+}(z) &=&\left( k-2\right) i\partial _zX^{+}e^{-2u/(k-2)}
\label{canonical} \\
S(z) &=&i\partial _zu  \nonumber
\end{eqnarray}
and then the currents take the form
\begin{eqnarray}
J^0(z)+J^1(z) &=&P^{+}(z) \\
J^2(z) &=&\,:X^{-}P^{+}:+S  \nonumber \\
J^0(z)-J^1(z) &=&\,:X^{-}P^{+}X^{-}:\,+2SX^{-}-ik\partial _zX^{-}  \nonumber
\end{eqnarray}
This is the form used in the previous section without the extra field $%
L^{\prime }(z).$ Thus, as discussed before, only the principal series will
emerge in the WZW model. Using the oscillator form introduced in (\ref
{fields}$)$ we can express $u(z)$ and $X^{+}(z)$ in terms of the basic
oscillators $s_n,\alpha _n^{+}$ by inverting the formulas in (\ref{canonical}%
), thus
\begin{eqnarray}
u(z) &=&u_0-is_0\ln z+i\sum_{n\neq 0}\frac 1ns_nz^{-n}  \label{ux} \\
X^{+}(z) &=&-i\int^zdz^{\prime }\frac{P^{+}(z^{\prime })}{\left( k-2\right) }%
\,\,:\exp \left[ \frac{2u(z^{\prime })}{k-2}\right] :\,.  \nonumber
\end{eqnarray}
Then these structures satisfy the operator products
\begin{equation}
\begin{array}{l}
<u(z)\,S(w)>=\left( \frac i{z-w}+\frac i{2w}\right) \left( \frac k2-1\right)
\\
\left[ J^0(z)-J^1(z)\right] \times X^{+}(w)
 \rightarrow \frac i{z-w}i:ie^{2u(w)/(k-2)}
\end{array}
\end{equation}
Thus, $u(z)$ is just the canonical conjugate to $S(z).$
Another property of $X^{+}$ that follows from the
fundamental operator products is that it is a singlet under the action of $%
J_2(z)$%
\begin{equation}
J_2(z)\times X^{+}(w)\rightarrow 0.  \label{j2xp}
\end{equation}
Actually $\partial X^{+}$ is a screening current. Its operator products with
all the currents is either zero or a total derivative. Therefore, its zero mode
commutes with all the currents.

Inserting these expressions in eq.(\ref{ux}) into (\ref{group}) we obtain
the quantum operator version of the group element $g$. The operartor
products may now be evaluated. We find the correct quantum products (\ref
{quantumrules}) with the above construction in terms of oscillators. That
is,
\begin{equation}
\begin{array}{l}
\left[ J^0(w)+J^1(w)\right] \times g(w)\rightarrow \frac{-i}{z-w}\left(
\begin{array}{ll}
0 & 0 \\
1 & 0
\end{array}
\right) g(w) \\
J^2(z)\times g(w)\rightarrow \frac{i/2}{z-w}\left(
\begin{array}{cc}
1 & 0 \\
0 & -1
\end{array}
\right) g(w) \\
\left[ J^0(w)-J^1(w)\right] \times g(w)\rightarrow \frac i{z-w}\left(
\begin{array}{ll}
0 & 1 \\
0 & 0
\end{array}
\right) g(w)
\end{array}
\end{equation}
This result, combined with the current $\times $ current operator products
that we have proven earlier, is convincing evidence that the free field
formalism that we have introduced corresponds to the quantization of the
SL(2,R) WZW model.

\section{Physical states}

\subsection{No ghosts}

Since we have rewritten the WZW theory in terms of free fields, the space of
states consists of the Fock space for the oscillators $\alpha _n^{\pm },s_n$
applied on the base $|p^{+},p^{-},s_0>$ that diagonalizes the zero mode
operators $\alpha _0^{\pm },s_0.$
\begin{equation}
\prod_{n=1}^\infty \left( \alpha _{-n}^{+}\right) ^{a_n}\prod_{m=1}^\infty
\left( \alpha _{-m}^{-}\right) ^{b_m}\prod_{k=1}^\infty \left( s_{-k}\right)
^{c_k}\,|p^{+},p^{-},s_0>
\end{equation}
where the powers $a_n,b_m,c_k$ are positive integers or zero. This is the
space of states that provide a representation basis for the SL(2,R) currents
with only the principal series. The physical states are identified as those
linear combinations that are annihilated by the the total Virasoro
generators
\begin{equation}
L_n|\psi >=0,\rm{ }\quad n\geq 1.
\end{equation}
In the present case the total Virasoro generators include the following
terms
\begin{equation}
L_n=L_n^{\pm }+L_n^S.
\end{equation}
where
\begin{eqnarray}
L_n^{\pm } &=&\sum_m:\alpha _{-m}^{-}\alpha _{n+m}^{+}: \\
L_n^S &=&\frac 1{k-2}\left( \sum_m\,:s_{-m}s_{n+m}:\,+ins_n+\frac 14\delta
_{n,0}\right)  \nonumber
\end{eqnarray}
Note that the $L_n^{\pm }$ is equivalent to the $c=2$ Virasoro operator in
2D flat spacetime$.$ The full central charge is
\begin{equation}
c=\frac{3k}{k-2}.
\end{equation}
The eigenvalue of the total $L_0$ is
\begin{equation}
L_0=p^{+}p^{-}+\frac 1{k-2}\left[ s_0^2+1/4\right] +%
\mathop{\rm integer}
\end{equation}
Thus, the theory has been reduced to a 2D lightcone in flat spacetime plus a
Liouville type space-like free field that has positive norm. A small but
important difference as compared to the standard Liouville formalism is that
the linear term in $L_n^S$ is Hermitean in our case, and does not contribute
to $L_0^S.$

The only negative norm states are the ones produced by the time-like
oscillator $\alpha _n^0=\left( \alpha _n^{+}-\alpha _n^{-}\right) /\sqrt{2}.$
However, this is no worse than the usual flat spacetime case\footnote{%
We have left out the $L^{\prime }$ sector since this is not part of the WZW
model. In a model that includes $L^{\prime }$ the base state would have the
additional label $h^{\prime }$ and products of Virasoro operators $%
L_{-n}^{\prime }$ applied on it. If the $L^{\prime }$ sector is present,
there could be additional negative norm states arising in sectors of
negative $h^{\prime },$ {\it if such values are permitted} by the model
describing $L^{\prime }$. For example, note that the norm of $L_{-n}^{\prime
}|h^{\prime }>$ is $2nh^{\prime }.$ Recall that negative values of $%
h^{\prime }$ lead to the discrete series when $j(j+1)>-1/4$. This
observation is, in fact, an explanation of the origin of negative norm
states that arose in the beginning.}. The space of physical states is
defined by
\[
\left( L_n-a\delta _{n,0}\right) |\phi >=0
\]
with $a\leq 1$ fixed. A proof of no ghosts can now be given by following
step by step the same arguments that prove the no ghost theorem in flat
spacetime \cite{noghost}. There is no need to repeat it here. We only recall
that there are no ghosts as long as $a\leq 1$ and $c\leq 26.$

It is straightforward to construct a few low lying states that satisfy the
Virasoro constraints and check explicitly that they have positive norm. For
example at level $l=1$ the mass shell condition is
\begin{equation}
p^{+}p^{-}+\frac 1{k-2}\left[ s_0^2+1/4\right] +1=a  \label{mshell}
\end{equation}
One finds the two orthogonal states that satisfy the Virasoro constraints
\begin{eqnarray}
&&
\begin{array}{l}
|\phi _1>=\left( p^{+}\alpha _{-1}^{-}-p^{-}\alpha _{-1}^{+}\right)
|p^{+},p^{-},s_0>
\end{array}
\nonumber \\
&&_{|\phi _2>=\left( (p^{+}\alpha _{-1}^{-}+p^{-}\alpha _{-1}^{+})(s_0+\frac
i2)-2p^{+}p^{-}s_{-1}\right) |p^{+},p^{-},s_0>}  \nonumber \\
&<&\phi _1|\phi _1>=-2p^{+}p^{-},\quad <\phi _1|\phi _2>=0, \\
&<&\phi _2|\phi _2>=-4p^{+}p^{-}(1-a).  \nonumber
\end{eqnarray}
After taking into account (\ref{mshell}) one sees that the norms are
positive as long as $a<1.$ The second state becomes a zero norm state if the
theory is critical, with $a=1.$ The first state survives even at the
critical point, and is interpreted as a string state in the 2D subspace. At
higher levels one finds that the number of states correspond to the same
counting as if there are two free bosons. This should be expected since we
started with three free bosons and basically eliminated one of them through
the Virasoro conditions. Some of the norms, but not all, are proportional to
$\left( 1-a\right) .$ If the theory is critical with $a=1$ a subset of these
states become zero norm states. The same phenomena can be observed in 3D
flat string theory when it is considered as a piece of the $D=26$ critical
theory. In fact, as $k\rightarrow \infty $ the states described above become
the states of the flat 3D non-critical theory, satisfying the 3D flat-string
Virasoro constraints. This is seen by rescaling $s_n\rightarrow \alpha _n^3%
\sqrt{k/2-1},$ where $\left[ \alpha _n^3,\alpha _m^3\right] =n\delta _{n+m}$
become the oscillators of the string in the third dimension, $\alpha
_0^3=s_0/\sqrt{k/2-1}$ becomes the momentum in the third dimension, and the $%
L_n$ of the present theory tend to the $L_n$ of the 3D flat string theory.
This observation is additional confirmation that the SL(2,R) curved space
theory behaves in the correct intuitive way as $k\rightarrow \infty $ by
becoming a flat space theory.

A more efficient approach to construct the physical states is to use the
``spectrum generating algebra'' as in flat space \cite{GSW}. This will be
presented in a separate paper.

\subsection{Monodromy}

So far we have not taken into account the physical effects of the $\ln z$
cut in the currents. As we argued in the beginning of section III, a
physical string theory must satisfy the monodromy condition in the physical
sector:
\begin{equation}
<phy|J^i(ze^{i2\pi n})|phy^{\prime }>=<phy|J^i(z)|phy^{\prime }>.
\label{monodrinv}
\end{equation}
This condition would have been satisfied automatically if there were no
cuts. One of the new features of our theory is to require that the monodromy
be satisfied only on a subset of states that are the physical states.
Quantum mechanically it is possible to impose this condition simultaneously
with the Virasoro constraints since the latter commute with the monodromy
operator as seen below.

To implement the monodromy let us first consider its effect on the currents.
{}From the modified currents in (\ref{newold}) we see that under the monodromy
the currents undergo a linear transformation

\begin{eqnarray}
\begin{array}{l}
\left[ J^0+J^1\right] (ze^{i2\pi n})=\left[ J^0+J^1\right] (z) \\
\left[ J^0-J^1\right] (ze^{i2\pi n})=\left[ J^0-J^1\right] (z)+4\pi n\alpha
_0^{-}\,\,J^2(z) \\
\quad \quad \quad \quad \quad \quad +\left( 2\pi n\alpha _0^{-}\right)
^2\left[ J^0+J^1\right] (z) \\
J^2(ze^{i2\pi n})=J^2(z)+2\pi n\alpha _0^{-}\,\left[ J^0+J^1\right] (z)
\end{array}
\label{monodr}
\end{eqnarray}
Therefore we expect that the right hand side can be rewritten as the adjoint
action with a {\it global} SL(2,R) transformation. Since the current $%
J^0(z)+J^1(z)$ remains unchanged the generator of this transformation must
be the zero mode of this current. Indeed, since $\alpha _0^{-}$ acts like a
number, we can rewrite the monodromy in the form
\begin{equation}
J^i(ze^{i2\pi n})=e^{-2i\pi n\alpha _0^{-}\left( J_0^0+J_0^1\right)
}J^i(z)\,e^{2i\pi n\alpha _0^{-}\left( J_0^0+J_0^1\right) }
\end{equation}
Therefore physical states that satisfy (\ref{monodrinv}) are the subset of
states that are invariant under the monodromy
\begin{equation}
e^{2i\pi n\alpha _0^{-}\left( J_0^0+J_0^1\right) }|phys>=|phys>.
\end{equation}
In the free boson representation this is easy to implement. Using $\left(
J_0^0+J_0^1\right) =\alpha _0^{+}\,$ this condition is applied on the Fock
space of the free bosons in the form
\begin{equation}
e^{2i\pi n\alpha _0^{-}\alpha _0^{+}}\prod_{n,m,k=1}^\infty \left( \alpha
_{-n}^{+}\right) ^{a_n}\left( \alpha _{-m}^{-}\right) ^{b_m}\left(
s_{-k}\right) ^{c_k}\,|p^{+}p^{-}s_0>
\end{equation}
Therefore it only requires that the momenta that describe the ground state
be quantized in terms of negative integers
\begin{equation}
\begin{array}{l}
e^{2i\pi n\alpha _0^{-}\alpha _0^{+}}|p^{+},p^{-},s_0>=|p^{+},p^{-},s_0> \\
\alpha _0^{-}\alpha _0^{+}=p^{-}p^{+}=-r,\quad r=0,1,2,\cdots
\end{array}
\end{equation}
We must take negative integers because according to the mass shell condition
$p^{-}p^{+}$ is negative. So, the mass shell condition on physical states at
excitation level $l$ takes the form
\begin{equation}
-r+\frac 1{k-2}\left[ s_0^2+1/4\right] +l=a.  \label{shell}
\end{equation}
It is always possible to satisfy this condition with some value of $s_0$
which is quantized in terms of the positive integers $r,l.$ In terms of the
original Casimir $j(j+1)$ this corresponds to a principal series
representation of SL(2,R) with quantized values of $j$ given by
\begin{eqnarray}
j &=&-\frac 12+is_0 \\
\ &=&-\frac 12\pm i\sqrt{\left( k-2\right) \left( r-l+a\right) -1/4}
\nonumber
\end{eqnarray}
where $r$ must be chosen so that the square root is real.

\subsection{Open and Closed strings}

An open string action $S=\int d\tau \int_0^\pi d\sigma \,L(\tau ,\sigma )$
is minimized by allowing free variation of the end points. For the WZW model
for any group $G$ this produces the boundary terms
\begin{equation}
\delta S=\int d\tau \,\left\{
\begin{array}{c}
\left. Tr\left( \left( \delta gg^{-1}\right) \left( \partial _\sigma
gg^{-1}\right) \right) \right| _\pi \\
-\left. Tr\left( \left( \delta gg^{-1}\right) \left( \partial _\sigma
gg^{-1}\right) \right) \right| _0
\end{array}
\right\}
\end{equation}
In addition to the equations of motion, these terms must also vanish at each
end of the string. That is,
\begin{equation}
\left. \partial _\sigma gg^{-1}\right| _{\sigma =0}=0=\left. \partial
_\sigma gg^{-1}\right| _{\sigma =\pi }.
\end{equation}
At the conformal critical point the equations of motion are satisfied by the
general form $g(\tau ,\sigma )=g_L(\tau +\sigma )\,g_R^{-1}(\tau -\sigma ).$
Then the boundary conditions require that $g_L$ and $g_R$ be related to each
other by the constraint
\begin{equation}
g_L^{-1}\left( \tau \right) \partial _\tau g_L(\tau )+g_R^{-1}\left( \tau
\right) \partial _\tau g_R(\tau )=0.  \label{relation}
\end{equation}
Furthermore, each term in this equation is required to be periodic. As
discussed in the rest of this paper, we impose periodicity on the physical
states. The relation (\ref{relation}) between $g_L(\tau )$ and $g_R(\tau )$
is not easy to solve explicitly. However, we may carry out the quantum
theory in terms of the current
\[
J(z)=g_L^{-1}\left( z\right) \partial _zg_L(z)=-g_R^{-1}\left( z\right)
\partial _zg_R(z).
\]
This is neither the left moving current $J_L=\partial g_Lg_L^{-1}$ nor the
right moving one $J_R=\partial g_Rg_R^{-1},$ but is related to them by
transformations involving $g_L$ or $g_R.$ This current generates
transformations on the {\it right side} of $g_L$ and the left side of $%
g_R^{-1},$ and the meaning of (\ref{relation}) is that the total current on
both $g_L$ and $g_R$ vanishes at the end points. The canonical commutation
rules for this current are identical to the ones we have already discussed
in the rest of the paper. The stress tensor constructed from it is equal to
the stress tensor constructed from either the left movers or the right
movers
\[
Tr(J^2)=Tr(J_L^2)=Tr(J_R^2).
\]
The quantum spectrum is obtained from the properties of $J$, whose
mathematical structure is the same as either left movers or right movers as
discussed in the previous sections. Thus, the quantum spectrum of the open
string in the SL(2,R) curved spacetime becomes identical to the spectrum
discussed above.

For a closed string we have independent left and right moving sectors. The
full group element is $g=g_L(z)g_R^{-1}(\bar z)$ and there are left and
right moving currents. Therefore we now need two sets of oscillators, the
left movers $\alpha _n^{\pm },s_n,$ and the right movers $\tilde \alpha
_n^{\pm },\tilde s_n.$ So, the direct product Hilbert space has a base
labelled by $|p^{-},p^{+},s_0;\tilde p^{-},\tilde p^{+},\tilde s_0>$ with $%
p^{-}p^{+}=-r$ and $\tilde p^{-}\tilde p^{+}=-\tilde r$ to insure that the
currents obey the monodromy conditions in the physical sector. We now need
to figure out if these are all independent labels or if they must be
constrained by physical considerations.

For this purpose we recall that a possible modular invariant is the so
called ``diagonal invariant'' that requires the same unitary representation
labelled by the same $j$ for both left and right movers. This may be
understood as being related to the representation of the full group element $%
D^j(g)=D^j(g_L(z))D^j(g_R^{-1}(\bar z))$ which requires the same $j$ for
both left and right movers. Therefore, we must demand $s_0=\tilde s_0.$

In addition, we examine $g(z,\bar z)$ in more detail. Keeping the order of
operators, it may be written in the form
\begin{equation}
g=g_L(z)g_R^{-1}(\bar z)=\left(
\begin{array}{cc}
u & a \\
-b & v
\end{array}
\right)
\end{equation}
with
\begin{eqnarray}
u &=&e^{\frac{-u_L+u_R}{k-2}}-e^{\frac{-u_L}{k-2}}\left(
X_L^{+}-X_R^{+}\right) X_R^{-}\,\,e^{\frac{-u_R}{k-2}} \\
v &=&e^{\frac{u_L-u_R}{k-2}}+e^{\frac{-u_L}{k-2}}X_L^{-}\left(
X_L^{+}-X_R^{+}\right) \,\,e^{\frac{-u_R}{k-2}}  \nonumber \\
a &=&e^{\frac{-u_L}{k-2}}\left( X_L^{+}-X_R^{+}\right) \,\,e^{\frac{-u_R}{k-2%
}}  \label{uvab} \\
b &=&-\left( X_L^{-}-X_R^{-}\right) e^{\frac{-u_L+u_R}{k-2}}  \nonumber \\
&&\quad +e^{\frac{-u_L}{k-2}}X_L^{-}\left( X_L^{+}-X_R^{+}\right)
X_R^{-}\,\,e^{\frac{-u_R}{k-2}}  \nonumber
\end{eqnarray}
We see that $g$ is not periodic under $\sigma \rightarrow \sigma +2\pi n$
since there are logarithms in the expressions for every $X_{L,R}^{\pm }$%
,\thinspace $u_{L,R}.$ However, provided we impose$\,$ $p^{+}=-\tilde p^{+}$
on physical states (to cancel the non-periodic behavior in $X_L^{+}-X_R^{+})$%
, we find that we can rewrite this monodromy in the form
\begin{eqnarray}
\begin{array}{l}
g(ze^{i2\pi n},\bar ze^{-i2\pi n})=U\tilde Ug(z,\bar z)\tilde U^{-1}U^{-1}
\\
U\tilde U=e^{-ip^{+}p^{-}2\pi n}e^{-is_0^22\pi n}\,e^{i\tilde p^{+}\tilde p%
^{-}2\pi n}e^{i\tilde s_0^22\pi n}
\end{array}
\end{eqnarray}
where $p^{+},s_0$ are operators which do not commute with $q^{-},u_0$, and
similarly for right movers (note that we have never introduced a canonical
conjugate to $p^{-}$ (or $\tilde p^{-})$). To insure that the matrix
elements of the overall $g$ are consistent with monodromy in the physical
sector it is sufficient to impose the conditions
\begin{equation}
2p^{+}p^{-}+2s_0^2-2\tilde p^{+}\tilde p^{-}-2\tilde s_0^2=2m
\end{equation}
where $m$ is an integer. Since we have already seen that $s_0=\tilde s_0$ we
find that this condition reduces to $r-\tilde r=m$, and does not impose any
additional constraints on $r,\tilde r.$

Furthermore, for a closed string we should also have $L_0-\tilde L_0=0$ on
the physical states. According to the mass shell condition (\ref{shell})
this requires $r-l=\tilde r-\tilde l.$ So, modular invariant physical closed
string states must be labelled at the base as follows
\begin{equation}
|-\frac r{p^{+}},\,\,p^{+},s>\times \,|\frac{\tilde r}{p^{+}},-p^{+},s>
\end{equation}
where the restrictions are
\begin{eqnarray}
\tilde p^{+} &=&-p^{+},\,\,\,\,\quad \tilde s_0=s_{0,} \\
p^{-}p^{+} &=&-r,\quad \tilde p^{-}\tilde p^{+}=-\tilde r  \nonumber
\end{eqnarray}
and the excitation numbers for left/right movers must be restricted by
\begin{equation}
r-l=\tilde r-\tilde l.
\end{equation}

\section{Comments}

Two novel features were introduced in this paper. The first is that currents
are allowed to contain logarithmic cuts provided monodromy conditions are
applied on the physical states. The second is a new representation of the
currents in terms of free bosons that render the theory completely solvable.
Both of these ideas have generalizations that would allow the construction
of a large number of new string models that are especially useful in curved
spacetime.

We have shown that a unitary string theory in SL(2,R) curved spacetime can
be constructed and its spectrum solved exactly. In a separate publication we
will give the spectrum generating algebra which characterizes the physical
states more efficiently. Correlation functions can also be computed by using
free boson methods.

Using the SL(2,R) solution given here, it is not difficult to figure out the
spectrum of the 2D black hole SL(2,R)/R gauged WZW model \cite{BN}\cite{WIT}%
. This requires imposing $J_n^2=\tilde J_n^2=0$ for $n\geq 1$ and $J_0^2+%
\tilde J_0^2=0$ on the SL(2,R) states described above. This task is more
easily carried out once the spectrum generating algebra is constructed. This
will be described elsewhere.

The new methods seem appropriate for understanding quantum string gravity
beyond the so called $c\geq 1$ barrier. In the present SL(2,R) case we have
solved a 3D model with $c=3k/(k-2)$ that can take values between $3$ and 26.

We have also shown that the free boson methods permit a more general
representation of SL(2,R) current algebra when the extra degrees of freedom $%
L_n^{\prime }$ are introduced. These were absent in the WZW model, but they
may be present in more general models.

As emphasized in the introduction, the main purpose for the present exercise
is to develop the appropriate methods to study string theory during the
early universe and to understand the impact of string theory on the
symmetries and matter content observed at accelerator energies. For this
purpose the current methods must be generalized to heterotic strings such as
those described in \cite{ibhetero}. Methods used for other special models of
curved spacetimes may also be helpful \cite{others}.

\section{Acknowledgements}

I thank J. Schulze and K. Pilch for helpful discussions.

\section{Appendix}

\subsection{Operator products-computation}

Here we compute the operator products of certain structures that will be
used as building blocks for the operator products of the currents given in
section V. The method of computation is to use Wick's theorem for free
fields to rearrange the oscillators into normal ordered form, then expand
the result in powers of the singularities, and finally drop the non-singular
terms (noted as $\cdots $ below). The following are needed in the
computation
\begin{equation}
\begin{array}{l}
P^{+}(z)\,\times \,:X^{-}(w)\,P^{+}(w): \\
\quad =:P^{+}(z)\,X^{-}(w)\,P^{+}(w): \\
\quad \quad +<P^{+}(z)\,X^{-}(w)>P^{+}(w) \\
\quad \rightarrow \frac{-\,i}{z-w}P^{+}(w)+\cdots
\end{array}
\label{pxqp}
\end{equation}
\begin{equation}
\begin{array}{l}
P^{+}(z)\times \,:X^{-}(w)\,P^{+}(w)X^{-}(w): \\
\quad =\,:P^{+}(z)\,X^{-}(w)\,P^{+}(w)X^{-}(w): \\
\quad \quad +2<P^{+}(z)\,X^{-}(w)>\,:X^{-}(w)\,P^{+}(w): \\
\quad \rightarrow \frac{-2\,i}{z-w}(:X^{-}(w)\,P^{+}(w):)+\cdots
\end{array}
\label{pxqpq}
\end{equation}

\begin{equation}
\begin{array}{l}
\left( :X^{-}\,P^{+}:\right) (z)\times \,\left( :X^{-}\,P^{+}:\right) (w)=
\\
=\,:X^{-}(z)\,P^{+}(z)X^{-}(w)\,P^{+}(w): \\
+\,:X^{-}(z)\,\,P^{+}(w):\,<P^{+}(z)X^{-}(w)>+ \\
+\,\,:P^{+}(z)\,X^{-}(w):\,<X^{-}(z)P^{+}(w)> \\
<P^{+}(z)X^{-}(w)>\,<X^{-}(z)P^{+}(w)> \\
\quad \rightarrow \frac 1{\left( z-w\right) ^2}+\cdots
\end{array}
\label{qpxqp}
\end{equation}
$\,$Similarly,
\begin{equation}
\begin{array}{l}
\left( :X^{-}\,P^{+}:\right) (z)\times \,\left( :X^{-}\,P^{+}X^{-}:\right)
(w)= \\
=:X^{-}(z)\,P^{+}(z)X^{-}(w)\,P^{+}(w)X^{-}(w): \\
\quad +2<P^{+}(z)X^{-}(w)>\,:X^{-}(z)\,\,P^{+}(w)X^{-}(w): \\
\quad +<X^{-}(z)\,P^{+}(w)>\,:\,X^{-}(w)\,P^{+}(z)X^{-}(w): \\
\quad +\,2<P^{+}(z)X^{-}(w)>\,<X^{-}(z)\,P^{+}(w)>X^{-}(w) \\
\quad \quad \rightarrow \frac{2X^{-}(w)}{\left( z-w\right) ^2}+\frac{%
-i\,\left( :X^{-}(w)\,\,P^{+}(w)X^{-}(w)\right) }{z-w}\,+\cdots
\end{array}
\label{qpxqpq}
\end{equation}
Furthermore,

\begin{equation}
\begin{array}{l}
_{\left( :X^{-}PX^{-}:\right) \left( z\right) \times \,\left(
:X^{-}\,PX^{-}:\right) \left( w\right) =} \\
=\,:\left( X^{-}\,PX^{-}\right) (z)\left( X^{-}\,P^{+}X^{-}\right) (w): \\
+2<X^{-}(z)P^{+}(w)>:\,P^{+}(z)X^{-}(z)X^{-}(w)\,X^{-}(w): \\
+2<P^{+}(z)X^{-}(w)>:\,X^{-}(z)X^{-}(z)X^{-}(w)\,P^{+}(w): \\
+4<X^{-}(z)P^{+}(w)><P^{+}(z)X^{-}(w)>:\,X^{-}(z)X^{-}(w): \\
\quad \quad \rightarrow \frac{4:\,X^{-}(z)X^{-}(w):}{\left( z-w\right) ^2}%
+\cdots \\
\quad \quad \rightarrow \frac{4:\,X^{-}(w)X^{-}(w):}{\left( z-w\right) ^2}+%
\frac{4:X^{-}(w)\partial _wX^{-}(w):}{z-w}+\cdots
\end{array}
\label{qpqxqpq}
\end{equation}
and
\begin{equation}
\begin{array}{l}
\left( :X^{-}(z)\,P^{+}(z)X^{-}(z):\right) \times \,\left( -ik\partial
_wX^{-}(w)\right) \\
\quad =-ik\,\,:X^{-}(z)\,P^{+}(z)X^{-}(z)\partial _wX^{-}(w): \\
\quad \,\,\,\,\,\,\,-ik\,\partial _w<P^{+}(z)X^{-}(w)>\,:X^{-}(z)X^{-}(z):
\\
\quad \rightarrow \,-k\frac{:X^{-}(z)X^{-}(z):}{(z-w)^2}+\cdots \\
\quad \rightarrow -k\frac{:X^{-}(w)X^{-}(w):}{(z-w)^2}-2k\frac{%
:X^{-}(w)\partial _wX^{-}(w):}{z-w}+\cdots
\end{array}
\label{qpqxdq}
\end{equation}
Similarly,
\begin{equation}
\begin{array}{c}
\left( -ik\partial _zX^{-}(z)\right) \times \left(
:X^{-}\,P^{+}X^{-}:\right) (w) \\
\rightarrow -k\frac{:X^{-}(w)X^{-}(w):}{(z-w)^2}+\cdots
\end{array}
\label{dqxqpq}
\end{equation}
Combining the last three equations we see that
\begin{equation}
\begin{array}{l}
_{\left[ :X^{-}\,P^{+}X^{-}:\,-ik\partial _zX^{-}\right] (z)\times \,\left[
:X^{-}P^{+}X^{-}:\,-ik\partial _wX^{-}\right] (w)\,} \\
\rightarrow 2(2-k)\left[
{\textstyle {:X^{-}X^{-}(w): \over (z-w)^2}}
+%
{\textstyle {:X^{-}\partial _wX^{-}: \over z-w}}
\right] +\cdots
\end{array}
\label{qpqdq}
\end{equation}
If $k$ had the value $2$ this operator product would not be singular.
However, for any $k$ the singularity cancels by including the current $S$ in
a modified operator as follows. Consider
\begin{equation}
\begin{array}{l}
\left[ 2X^{-}(z)S(z)\right] \times \,\left[ 2X^{-}(w)S(w)\right] \\
\quad =\,\,4\,\,\left( :X^{-}(z)X^{-}(w):\right) \,\,\left( :S(z)S(w):\right)
\\
\quad \quad +4\frac{k/2-1}{\left( z-w\right) ^2}\left(
:X^{-}(z)X^{-}(w):\right) \\
\rightarrow 2(k-2)\left[
{\textstyle {:X^{-}(w)X^{-}(w): \over (z-w)^2}}
+%
{\textstyle {:X^{-}(w)\partial _wX^{-}(w): \over z-w}}
\right] +\cdots
\end{array}
\label{qsxqs}
\end{equation}
and the combination
\begin{equation}
{\left[ :X^{-}\,P^{+}X^{-}:\,-ik\partial _zX^{-}\right] (z)\times \left[
2X^{-}S\right] (w) \atopwithdelims\{\} +\left[ 2X^{-}S\right] (z)\times \left[
:X^{-}\,P^{+}X^{-}:\,-ik\partial _wX^{-}\right] (w)}
\rightarrow 0  \label{qpqdqxqs}
\end{equation}
which is not singular (although each term by itself is). Combining the last
three equations we see that the following operator product is not singular
\begin{equation}
_{
\begin{array}{l}
\left[ :X^{-}\,P^{+}X^{-}:\,-ik\partial _zX^{-}+2X^{-}S\right] (z)\times \\
\quad \times \,\left[ \,:X^{-}\,P^{+}X^{-}:\,-ik\partial
_wX^{-}+2X^{-}S\right] (w) \\
\quad \quad \quad \quad \rightarrow \,\,\,0+\cdots
\end{array}
}  \label{qpqdqqs}
\end{equation}

\subsection{Operator products with $1/P^{+}$}

If the $L_n^{\prime }$ are included in the construction then we need to
compute the operator products with $1/P^{+}(z).$ This operator may be
treated as a series in the oscillators, with the zeroth order term $1/p^{+}.$
Successive terms in the series contain higher powers of $1/p^{+}.$ The
series is well defined provided one acts on states for which $1/p^{+}$ is a
well defined operator. In momentum space $|p^{+}>$ this simply requires a
non-zero eigenvalue $p^{+}\neq 0.$ In the space
\mbox{$\vert$}
$j,m>$ labelled by the eigenvalues of $J_0$ one has to be careful since the
behavior of the wavefunction $<p^{+}|j,m>$ near the origin is ($%
p^{+})^{j+1/2+is}$ (discrete,supplementary, principal series) or ($%
p^{+})^{-j-1/2-is}$ (principal, supplementary series). Sufficiently high
powers of $1/p^{+}$ may map a given state $|jm>$ out of the normalizable
Hilbert space. This would have to be interpreted properly. In the following
we assume that the operators are well defined on an appropriate set of
states, such as the states $|p^{+}\neq 0>$. Then
\begin{equation}
\begin{array}{l}
X^{-}(z)\frac 1{P^{+}(w)}=\,:X^{-}(z)\frac 1{P^{+}(w)}:\, \\
\quad \quad \quad \quad \quad \quad -\left( \frac 1{P^{+}(w)}\right)
^2<X^{-}(z)P^{+}(w)> \\
\quad \quad \quad \quad \rightarrow \frac{-i}{\left( z-w\right) }\left(
\frac 1{P^{+}(w)}\right) ^2+\cdots
\end{array}
\label{q/p}
\end{equation}
This leads to
\begin{equation}
\begin{array}{l}
:X^{-}(z)\,P^{+}(z):\times \frac 1{P^{+}(w)}\rightarrow \,\frac{-i}{\left(
z-w\right) }\frac 1{P^{+}(w)}+\cdots \\
\frac 1{P^{+}(z)}\times :X^{-}(w)\,P^{+}(w):\rightarrow \frac i{\left(
z-w\right) }\frac 1{P^{+}(w)}+\cdots \\
:X^{-}\,P^{+}X^{-}:(z)\,\times \frac 1{P^{+}(w)}\rightarrow \,\frac{-2i}{%
\left( z-w\right) }\,:X^{-}(w)\frac 1{P^{+}(w)}: \\
\quad \quad \quad \quad \quad \quad \quad \quad \quad \quad \quad \quad +%
\frac{-2}{(z-w)^2}\frac 1{\left( P^{+}(w)\right) ^2}+\cdots
\end{array}
\end{equation}
and
\begin{equation}
\begin{array}{l}
\left[ :X^{-}\,P^{+}X^{-}:\,-ik\partial _zX^{-}+2X^{-}S\right] (z)\,\times
\frac{T^{\prime }(w)}{P^{+}(w)} \\
\rightarrow \,\quad \,\frac{-2iT^{\prime }(w)}{\left( z-w\right) }\left(
:X^{-}\frac 1{P^{+}}:\,+S\right) (w) \\
\quad \quad \quad +\frac{k-2}{(z-w)^2}\frac{T^{\prime }(w)}{\left(
P^{+}(w)\right) ^2}+\cdots
\end{array}
\end{equation}
Similarly,
\begin{equation}
\begin{array}{l}
\frac{T^{\prime }(z)}{P^{+}(z)}\times \,\left[
:X^{-}\,P^{+}X^{-}:\,-ik\partial _wX^{-}+2X^{-}S\right] (w) \\
\rightarrow \frac{2iT^{\prime }(w)}{\left( z-w\right) }\left( :X^{-}\frac 1{%
P^{+}}:\,+S\right) (w)+\frac{k-2}{(z-w)^2}\frac{T^{\prime }(w)}{\left(
P^{+}(w)\right) ^2} \\
\quad \quad \quad +\,\frac{k-2}{(z-w)}\partial _w\left( \frac{T^{\prime }(w)%
}{\left( P^{+}(w)\right) ^2}\right) +\cdots
\end{array}
\label{ts/qpqdqqs}
\end{equation}
We also have
\begin{equation}
\begin{array}{l}
\frac{T^{\prime }(z)}{P^{+}(z)}\times \frac{T^{\prime }(w)}{P^{+}(w)}%
\rightarrow \frac 1{P^{+}(z)P^{+}(w)}\left[ \frac{2T^{\prime }(w)}{(z-w)^2}+%
\frac{\partial _wT^{\prime }(w)}{\left( z-w\right) }\right] \\
\quad \quad \rightarrow \frac 1{(z-w)^2}\frac{T^{\prime }(w)}{\left(
P^{+}(w)\right) ^2}+\frac 1{(z-w)}\partial _w\left( \frac{T^{\prime }(w)}{%
\left( P^{+}(w)\right) ^2}\right) +\cdots
\end{array}
\label{t/pxt/p}
\end{equation}
Combining the last three equations together with (\ref{qpqdqqs}) shows that
the current

\begin{equation}
\begin{array}{l}
J_0(z)-J_1(z)=:X^{-}(z)\,P^{+}(z)X^{-}(z): \\
\,\,\,\,-ik\partial _zX^{-}(z)+2X^{-}(z)S(z)-%
{\textstyle {(k-2)T^{\prime }(z) \over P^{+}(z)}}
\end{array}
\,
\end{equation}
has non-singular operator product with itself
\begin{equation}
\left( J_0(z)-J_1(z)\right) \times \left( J_0(z)-J_1(z)\right) \rightarrow
\,0+\cdots  \label{j01xj01}
\end{equation}


\end{document}